\documentclass[12pt]{iopart}
\usepackage{amsmath}
\usepackage{amssymb}
\usepackage{amsthm}
\usepackage{amstext}
\usepackage{array}
\usepackage{xypic}
\usepackage{bm}
\usepackage[T1]{fontenc}
\usepackage[cp1250]{inputenc}

\eqnobysec

\newcommand{\AB}{\allowbreak}

\newcommand{\ali}[2]{\mathop{\mathfrak{#1}(#2)}\nolimits}

\newcommand{\ad}{\mathop{\mathrm{ad}}\nolimits}
\newcommand{\ADA}[1]{\ifmmode \ad(#1) \else $\ad(#1)$\fi}
\newcommand{\LI}[2]{\ifmmode#2_1,\AB\,\ldots,\,\AB #2_{#1}%
\else$ #2_1,\AB\,\ldots,\,\AB#2_{#1}$\fi}

\newcommand{\su}[1]{\ali{su}{#1}}
\newcommand{\sltwo}{\ifmmode \ali{sl}{2} \else $\ali{sl}{2}$\fi}

\long\def\comment#1{}

\newcommand{\bMA}[1]{\[\begin{array}{#1}}
\newcommand{\eMA}{\end{array}\]}

\newcommand{\C}{{{\mathbb C}}}

\newcommand{\NR}{{\mathbb{R}}}

\def\be{\begin{equation}}
\def\ee{\end{equation}}

\def\R{\NR}

\def\cpn{{\C P^{N-1}}}
\def\bp{{\bar{\partial}}}
\def\p{{\partial}}

\def\tr{{\mathrm{tr}}}

\newcommand{\bea}{\begin{eqnarray}}
\newcommand{\eea}{\end{eqnarray}}
\newcommand{\la}{\lambda}
\renewcommand{\c}{\cdot}

\begin{document}

\title{Invariant formulation of surfaces associated with $\cpn$ models}

\author{P. P. Goldstein$^1$ and A. M. Grundland$^2$}

\address{$^1$ Theoretical Physics Department, \\
The Andrzej Soltan Institute for Nuclear Studies, \\
Hoza 69, 00-681 Warsaw, Poland}
\ead{Piotr.Goldstein@fuw.edu.pl}
\vspace{5mm}
\address{$^2$ Centre de Recherches Math{\'e}matiques, Universit{\'e} de Montr{\'e}al, \\
C. P. 6128, Succ.\ Centre-ville, Montr{\'e}al, (QC) H3C 3J7, Canada \\
Universit\'{e} du Qu\'{e}bec, Trois-Rivi\`{e}res CP500 (QC) G9A 5H7, Canada}
\ead{grundlan@crm.umontreal.ca}


\begin{abstract}
In this paper, we provide an invariant formulation of completely
integrable $\cpn$ Euclidean sigma models in two dimensions defined on
the Riemann sphere $S^2$. The scaling invariance is explicitly taken
into account by expressing all the equations in terms of projection
operators. Properties of the projectors mapping onto one-dimensional
subspaces are discussed in detail. The paper includes a discussion of
surfaces connected with the $\cpn$ models and the wave functions of
their linear spectral problem.
\end{abstract}

\maketitle

\section{Introduction}
\label{sec:Intro} The simple Dirichlet Lagrangian
\be
\mathcal{L} = \frac{1}{4}(D_{\mu}z)^{\dagger}\cdot(D_{\mu}z),
\label{lagr-z}
\ee
where $^\dagger$ denotes the Hermitian conjugate, may represent e.g.
a free quantum particle described by a wave function $z$. It is
trivial when $z$ is a scalar function of its variables and $D_{\mu}$
is just a partial derivative. However the model becomes nontrivial
and has found many applications if the target space is a complex
Grassmanian manifold and the partial derivatives $(D_{\mu}z)$ turn
into the appropriate covariant derivatives. The most popular are
$\cpn$ models, whose target spaces are complex $G(1,N)$ Grassmanians,
equivalent to projective spaces ($\C P$ stands for complex
projective). The target space is a set of lines intersecting at the
origin or, equivalently, the $N-1$ dimensional Riemann sphere immersed
in an $N$-dimensional vector space. The two-dimensional space of
independent variables is the simplest nontrivial one. In that case
the variables $z$ are subject to the constraint
\be\label{constraint}
z^{\dagger}\cdot z=1,
\ee
and the covariant derivative has the form
\be
\label{cov}
D_{\mu}z=\partial_{\mu}z-(z^{\dagger}\cdot\partial_{\mu}z)z,\qquad\partial_{\mu}
=\partial_{\xi^{\mu}},\qquad\mu=1,2.
\ee

These systems are the subject of our investigation in this paper.

The main objective of the paper is to formulate differential
projective geometry in terms of projective operators, which makes it
explicitly invariant under scaling by any scalar function. Some
applications of the projectors have been introduced earlier in
\cite{Mik,WZ}. In this paper we construct a basis of projectors which
map onto orthogonal one-dimensional subspaces and use it to express all
other quantities. As the model is exactly solvable \cite{Mik,MikZ},
the formulation encompasses the spectral problem and the surfaces
whose immersion conditions are the dynamics equation of the system.

Instead of the Cartesian variables $(\xi^1,\xi^2)\in \R^2$, we use
more convenient complex variables $(\xi,\bar{\xi})~\in \C$, where
$\xi=\xi^1+i\,\xi^2$ (complex conjugates are marked by a bar over a
symbol). The complex plane is usually compactified to the Riemann
sphere.

To avoid the inconvenient non-analytic condition
\eqref{constraint} the model dynamics is usually expressed in
terms of
\be
z=f/|f|,\quad |f|=\left(f^{\dagger}\cdot f\right)^{1/2},
\ee
without any constraints on the new variable $f$. The Euler-Lagrange
(E-L) equations in the new variables read
\be
\label{E-L} \left(\mathbb{I}-\frac{f\otimes
f^{\dagger}}{f^{\dagger}\cdot f}\right)\cdot\left[\p\bp
f-\frac{1}{f^{\dagger}\cdot f}\left((f^{\dagger}\cdot\bp f)\p
f+(f^{\dagger}\cdot\p f)\bp f\right)\right]=0,
\ee
where $\p$ and $\bp$ are derivatives with respect to the complex
independent variables $\xi$ and $\bar{\xi}$ respectively,
$\mathbb{I}$ is the $N\times N$ unit matrix.

On the other hand, equation \eqref{E-L} does not have the simplicity
of the original Lagrangian. Moreover the solutions in terms of the
new variables are not unique: the same Grassmanian solution
corresponds to infinitely many $f$'s. To achieve uniqueness, a
constraint has to be imposed on $f$; most often it relies on putting
its first nonzero component equal to $1$.

It  seems  that more natural variables in the $\cpn$ and all
Grassmanian models are projection operators, more precisely
orthogonal projectors mapping onto individual directions in the
$\cpn$ models or on the appropriate subspaces in Grassmanians of
higher order. The orthogonal projector which maps onto a
one-dimensional subspace in the direction $f$ may be written as
\be
P=[1/(f^\dagger\!\c\! f)]f\otimes f^{\dagger}.
\ee
It is evident that such projectors (as well as other orthogonal
projectors) are Hermitian $P^\dagger=P$. They are also subject to
a constraint, but the constraint is analytic and simple
\be\label{Pconstraint}
P^2=P,
\ee
while the Lagrangian is as simple as the one for $z$
\cite{WZ,Manton}:
\be
\label{action2} \mathcal{L}=\tr(\p P\cdot\bp P).
\ee
Similarly to the case of the $z$ variables, the appropriate constraint,
\eqref{Pconstraint} in this case, is multiplied by a Lagrange
multiplier and subtracted from the Lagrangian before taking the
variation of the action integral. The variation yields the E-L
equations for the projectors, which can be expressed in the well-known form
of a conservation law \cite{MikZ, WZ, Manton}, namely
\be\label{cons-law}
\p\,[\bp P,P]+\bp\,[\p P,P]=0.
\ee
More details on the $\cpn$ sigma models may be found in
\cite{WZ,Manton}. In the present paper we concentrate on the consequences
of expressing their theory in terms of the projectors.

 This paper is organized as follows. In section 2 we list the basic
 algebraic and analytical properties of orthogonal projectors which map onto
 one-dimensional  subspaces. Invariant recurrence relations, which follow from
 those properties, are summarized without much detail (the detailed
 discussion is given in our other paper \cite{GG1}). In section 3
 we discuss the mutual connection between the projectors and the
 surfaces whose conditions for immersion (the Gauss-Codazzi-Ricci
 equations) are equivalent to \eqref{cons-law}. Section 4 contains
 a discussion of the linear problem and the corresponding wave
 function. Finally, we list the conclusions and possible directions for
 further work.

\section{Properties of projectors mapping onto $1^D$
subspaces}

Here we list the properties of the orthogonal projection matrices $P$
which map onto one-dimensional subspaces. All of the discussed properties
follow from the defining property \eqref{Pconstraint} and from the
fact that the target is one-dimensional.
\begin{enumerate}
\item
From the definition of the projective property \eqref{Pconstraint}
it follows that the operators $P$ are diagonalizable and their
eigenvalues are $0$ or $1$.
\item
If a projector maps onto a one-dimensional subspace, its rank is one
and thus only one of the eigenvalues is one, the rest being zero.
Hence for such $P$
\be
\tr (P) = 1.
\ee
The diagonalisation may always include placing the only nonzero
eigenvalue in the first row and first column.
\item
By differentiating the defining property \eqref{Pconstraint} we
obtain after straightforward computation
\be\label{exch-P}
\p P\cdot P = (\mathbb{I}-P)\c\p P,\quad P\cdot\p P = \p
P\cdot(\mathbb{I}-P)
\ee
and the same holds for the ``barred'' derivative $\bp$. In other
words: an exchange with $\p P$ or $\bp P$ turns $P$ into
$\mathbb{I}-P$ and vice versa.

Induction yields a more general property about the exchange of $P$
with an arbitrary number of $\p P$ and $\bp P$ in arbitrary order
\be\label{even-P}
P\cdot \p P\cdot\bp P\cdot...\cdot\p P = \p P\cdot\bp
P\cdot...\cdot\p P\cdot P
\ee
if the total number of the derivatives $\p P,~\bp P$ is even, or
\be\label{poly-P}
P\cdot \p P\cdot\bp P\cdot...\cdot\p P = \p P\cdot\bp
P\cdot...\cdot\p P\cdot(\mathbb{I}-P)
\ee
if the total number of the derivatives $\p P,~\bp P$ is odd.

The above properties hold for all projection operators, regardless of
the dimension of their target subspace and the projection angle.
\item
If an orthogonal projector $P$ maps onto a one-dimensional subspace
then, for any square matrix $A$ having the same dimension as the space, we have
\be\label{tr-P}
P\cdot A\cdot P=\tr (P\cdot A)\,P.
\ee
The proof by diagonalisation follows directly from the property that
only one eigenvalue of $P$ is one, while the others are zero. A
consequence of this property is the necessary and sufficient
condition that a projection of any projector $Q$ onto the projector
$P$ is a zero matrix, that is
\be
P\c Q\c P= 0\quad \mathrm{iff}\quad \tr(P\c Q)=0,
\ee
which is compatible with the definition of the scalar product
\cite{GG1}
\be
(A,B)=-(1/2)\tr(A\c B).
\ee
\item
The following traces vanish:
\be\label{vanishP}
\tr(P\cdot\p P\c P\c\bp P\c ... \c\p P)=0,
\ee
where the matrix product of derivatives contains any odd number of
the $\p$ and $\bp$ derivatives in arbitrary order, while the
number of the projectors $P$ and their positions in the product
are also arbitrary.\\

\textit{Proof:} If the product contains at least one operator $P$:
Write any of the operators $P$ in the product as $P\c P$ and exchange
the right $P$ with $\p P$ and $\bp P$, one by one, up to the
rightmost position, while moving the left $P$ in the same way to the
leftmost position. On each exchange, $P$ turns into $\mathbb{I}-P$
and vice versa (property \ref{poly-P}). If we do not encounter the
product $P\c (\mathbb{I}-P)=0$ or $(\mathbb{I}-P)\c P=0$ (which ends
the procedure), we end up with
\be
\tr\left(P\cdot\p P\c P\c\bp P\c ... \c\p P\c
(\mathbb{I}-P)\right)
\ee
if the number of the derivatives to the left of the $P$ is even
while that to the right is odd,
 or
\be
\tr\left((\mathbb{I}-P)\cdot\p P\c P\c\bp P\c ... \c\p P\right)
\ee
if the number of the derivatives to the left of the $P$ is odd while
that to the right is even. In either case the trace is zero because a
cyclic permutation of the factors yields an expression containing the
product $P\c (\mathbb{I}-P)=0$ or $(\mathbb{I}-P)\c P=0$.

If the product contains only derivatives, without $P$ operators, the
unit matrix may be put as the first factor and represented as
$P+(\mathbb{I}-P)$. The same procedure as before yields zero for
each of the components. \textit{Q.E.D.}
\item
Properties involving $2^\mathrm{nd}$ derivatives of the projectors
$P$ are interesting as such derivatives occur in the E-L equations
\eqref{cons-law}. It follows from $\tr(P\cdot\p P)=0$ and
$\tr(P\cdot\bp P)=0$ (a special case of \eqref{vanishP}) that
\be\label{2-d deriv}
\tr(P\c\p ^2 P)=-\tr(\p P\c\p P),
\ee
with analogous formulae for the $\p\bp$ and $\bp ^2$ derivatives.
\item
If the $P$ operator also satisfies the E-L equations
\eqref{cons-law}, then we have
\be
\tr(P\c\bp P\c \p\bp P)=0
\ee
The proof is straightforward if we use property \eqref{exch-P} and
the invariance of traces on cyclic permutations.
\item
While traces of products of an odd number of derivatives vanish for
the $P$ projectors, the traces of an even number can significantly be
simplified by the following property: For any square matrix $A$ (of
the proper dimension) we have the factorisation property
\be\label{factor-P}
\tr(A\c\p P\c\p P\c P)=\tr(A\c P)\tr(\p P\c\p P\c P),
\ee
with analogous formulae in which one or both $\p$ derivatives are
replaced by $\bp$.\\

\textit{Proof:} From equation \eqref{even-P} we have
\be
\tr(A\c\p P\c\p P\c P)=\tr(A\c P\c\p P\c\p P\c P),
\ee
which yields equation \eqref{factor-P} by property \eqref{tr-P}.

\item\textbf{Gram-Schmidt orthogonalization}\\
Given a set of linearly independent vectors in a vector space, we
can always construct an orthonormal basis by using the well-known
Gram-Schmidt orthogonalization procedure. An orthogonal basis has
its counterpart in the corresponding set of projectors. If we
represent one-dimensional projectors as matrices in the new
orthonormal basis, the $i$-th projector $P_i$ is represented by a
matrix having one nonzero diagonal element, $z_{ii}=1$, while all
other elements of the matrix are zero.

This straightforward procedure is not so trivial if the vectors and
projectors are functions of the mainfold parameters e.g.
$\xi,\bar{\xi}$ and we want the basis vectors to satisfy the E-L
equations \eqref{E-L} or, equivalently, the corresponding projectors
$P_i$ to satisfy equation \eqref{cons-law}. The Gram-Schmidt
orthogonalization operators which map solutions of \eqref{E-L} to
consecutive orthogonal solutions are \cite{Din, WZ}
\be
\label{P+} P_+(f)=(\mathbb{I}-P)\c \p f,
\ee
which we refer to as a ``creation operator'' and
\be\label{P-}
P_-(f)=(\mathbb{I}-P)\c\bp f,
\ee
which is called an ``annihilation operator''. The corresponding
``creation'' and ``annihilation'' operators for one-dimensional
projectors were found in \cite{GG1}. They are defined by
\be
\label{Ppm} \mathbf{\Pi_-}(P_k)=P_{k-1}, \qquad
\mathbf{\Pi_+}(P_k)=P_{k+1}.
\ee
while their explicit form reads
\be
\label{Pi-} \mathbf{\Pi_-}(P)=\frac{\bp P\c P\c \p P}{\tr(\bp P\c
P \c\p P)}=\frac{(\mathbb{I}-P)\c \bp P\c \p P}{\tr(\bp P\c P\c\p
P)}=\frac{\bp P\c \p P\c (\mathbb{I}-P)}{\tr(\bp P\c P\c \p P)},
\ee
and
\be
\label{Pi+} \mathbf{\Pi_+}(P)=\frac{\p P\c P \c\bp P}{\tr(\p P\c P
\c\bp P)}=\frac{(\mathbb{I}-P)\c \p P\c \bp P}{\tr(\p P\c P\c\bp
P)}=\frac{\p P\c \bp P\c (\mathbb{I}-P)}{\tr(\p P\c P\c \bp P)},
\ee
 The complete basis is obtained if we apply the
creation operator $0,1,\ldots,N-1$ times to any holomorphic solution
of \eqref{E-L} or its projector counterpart \eqref{Ppm}. The
construction may also be performed in the opposite direction by means
of the annihilation operator, starting from an antiholomorphic
solution.

It immediately follows from (\ref{Pi-}, \ref{Pi+}), that the result
of the ``creation'' or ``annihilation'' is always orthogonal to the
original projector
\be
\mathbf{\Pi_+}(P)\c P=P\c\mathbf{\Pi_+}(P)=0 \quad \text{and}\quad
\mathbf{\Pi_-}(P)\c P=P\c\mathbf{\Pi_-}(P)=0
\ee
\item
If the basis is built by the above Gram-Schmidt orthogonalisation,
starting from a vector which is a holomorphic function of $\xi$, then
the projectors whose target subspaces are vectors of the basis,
satisfy
\be\label{dPdP}
\tr(\p P\c\p P)=0,\quad\tr(\bp P\c\bp P)=0.
\ee
\end{enumerate}
Other projectors mapping onto one-dimensional subspaces do not
have to satisfy this equation. The proof will be given in
Appendix A.

Using properties (i)--(x) we can prove all the properties required
for the model to be consistent.
\begin{enumerate}
\item
If $P$ is an orthogonal projector and $P_{+1}=\mathbf{\Pi_+}(P)$
exists, then $P_{+1}$ is also an orthogonal projector: it has the
projective property $P_{+1}^2=P_{+1}$ and its kernel is orthogonal to
its target subspace. The same is true of $\mathbf{\Pi_-}(P)$. Moreover
the trace of $\p P\c P\c \bp P$ vanishes iff the whole matrix
vanishes (the same holds for $\bp P\c P\c \p P$), which ensures the
possibility of constructing $P_{\pm 1}$ whenever the matrix is
nonzero.
\item
The operators $\mathbf{\Pi_+}$ and $\mathbf{\Pi_-}$ are inverses of
each other, i.e.
$\mathbf{\Pi_+}\left(\mathbf{\Pi_-}(P)\right)=\mathbf{\Pi_-}\left(\mathbf{\Pi_+}(P)\right)=P$,
provided that the inner operation on $P$ is possible.
\item
If $P$ satisfies the E-L equations \eqref{cons-law} and
$P_{+1}=\mathbf{\Pi_+}(P)$ exists, then $P_{+1}$ also satisfies those
equations.
\end{enumerate}
 The proofs are in Appendix A.

\section{Projectors and soliton surfaces}
We first recall some of the previously known results. It has been
shown in \cite{GSZ} that the conservation law \eqref{cons-law} may
be interpreted as a condition for the contour integral
\be\label{X}
X(\xi,\bar{\xi})=i\int_{\gamma}\left(-[\p P,P]d\xi+[\bp
P,P]d\bar{\xi}\right),
\ee
to be independent of the path of integration $\gamma$. This defines a
mapping of an area on a Riemann sphere into a set of $su(N)$ matrices
$\Omega\ni(\xi,\bar{\xi})\mapsto X(\xi,\bar{\xi})\in
su(N)\simeq\mathbb{R}^{N^2-1}$. This generalised Weierstrass formula
for immersion of 2D surfaces in $\mathbb{R}^{N^2-1}$
\cite{Kono1,Kono2,NS} defines surfaces in terms of the projectors
$P$. The compatibility conditions of the immersion constitute the
conservation law \eqref{cons-law}. The integration may be performed
explicitly for the surfaces corresponding to the projectors $P_k$
obtained recursively from the holomorphic solution. It yields
(\cite{GY1}, see also the proof in Appendix A)
\be
\label{XfromP}
X_k=-i\left(P_k+2\sum\limits_{j=0}^{k-1}P_j\right)+\frac{i(1+2k)}{N}\,\mathbb{I},
\qquad k=0,\ldots,N-2.
\ee
For $k=N-1$ equation \eqref{XfromP} gives an equation equivalent to
that for $k=0$, which reduces the number of surfaces (or
algebraically independent immersion functions).

Inversely, we can obtain the projectors $P_k$ from the surfaces $X_k$
either as a linear combination of the surfaces $X_0,\ldots , X_k$
\be
\label{PkX}
P_k=i\sum\limits_{j=1}^k
(-1)^{k-j}\left(X_j-X_{j-1}\right)+(-1)^k i
X_0+\frac{1}{N}\mathbb{I},
\ee
or by a nonlinear formula which depends on $X_k$ only \cite{GG1}
\be\label{Pk from Xk}
P_k={X_k}^2-2i\left(\frac{2k+1}{N}-1\right)X_k-\frac{2k+1}{N}
\left(\frac{2k+1}{N}-2\right)\mathbb{I}.
\ee

The projective property $P_k^2=P_k$ apparently imposes a constraint
on the surfaces $X_k$. Does it constitute an equation defining those
surfaces or is it identically satisfied by the surfaces constructed
from \eqref{XfromP}? To verify this, we examine the projective property
for $P_k$, where we substitute the $P_k$ with \eqref{Pk from Xk}.
Direct substitution of \eqref{Pk from Xk} into the projective property
yields a 4$^\mathrm{th}$ degree equation. However a simpler,
3$^\mathrm{rd}$ degree condition may be obtained by multiplying
\eqref{Pk from Xk} by $X_k$ and making use of the fact that
$P_k$ is orthogonal to all the lower-index projectors. The 3$^\mathrm{rd}$ degree
condition obtained in this way may be factorised to the form
\be\label{3-deg-cond}
\left[X_k-i\left(\frac{1+2
k}{N}-2\right)\mathbb{I}\right]\left[X_k-i\left(\frac{1+2
k}{N}-1\right)\mathbb{I}\right]\left[X_k-i\frac{1+2
k}{N}\mathbb{I}\right]=0.
\ee
This condition has a simple interpretation if we diagonalise it,
which is always possible as the $X_k$ matrices are antihermitian. The
diagonalised form \eqref{3-deg-cond} consists of a product of
matrices containing merely eigenvalues of $X_k$ minus a number equal
to $i[(1+2k)/N-2]$, $i[(1+2k)/N-1]$ or $i(1+2k)/N$. We find that
equation \eqref{3-deg-cond} is always satisfied if the surfaces have
been constructed according to \eqref{XfromP}. That is
\begin{itemize}
\item
The component $i(1+2k)/N$ has been added to each diagonal element of
the sum of projectors \eqref{XfromP} to make $X_k$ traceless.
Therefore it occurs as a component of every eigenvalue of $X_k$.
\item
The component $2i$ subtracted from $i(1+2k)/N$ is a contribution due
to $\-2i \sum_{j=0}^{k-1} P_j$ as each of the $P_j$ has one
eigenvalue equal to 1 and the other eigenvalues equal to 0. It occurs at
the indices of the dimensions onto which $P_1\ldots P_{k-1}$ map.
\item
The component $i$ subtracted from $i(1+2k)/N$ is a contribution
due to $-i\,P_k$. It is a component of the eigenvalue at the index
pointing at the dimension onto which $P_k$ maps.
\item
Nothing is subtracted from $i(1+2k)/N$ at the indices $k+1\ldots N$
pointing at the dimensions onto which none of $P_0,\ldots, P_k$ map.
\end{itemize}
Thus equation \eqref{3-deg-cond} is the lowest degree constraint on
the immersion functions $X_k$ of the surfaces. If we directly
substituted \eqref{Pk from Xk} into the projective property, we would
get an equivalent condition: the equation would differ from
\eqref{3-deg-cond} by the middle factor: in the $4^\mathrm{th}$
degree condition the factor
$\left[X_k-i\left((1+2k)/N-1\right)\mathbb{I}\right]$ is squared.

Although equation \eqref{3-deg-cond} is obvious when we look at the
source of $X_k$ \eqref{XfromP}, it is nevertheless a nontrivial
constraint on the surfaces. Since all the eigenvalues are independent
of the coordinates $(\xi, \bar{\xi})$, the whole kinematics of a
moving frame (vielbein) may only be due to variation of the
diagonalising (unitary) matrix.\\

\textbf{Differential geometry of the surfaces.} Once we have the immersion
functions of the surfaces, we can describe their metric and curvature
properties.
\begin{enumerate}
\item
The diagonal elements of the metric tensor are zero. This property,
proven in \cite{GG1}, directly follows from property \eqref{dPdP}.
Let $g_k$ be the metric tensor corresponding to the surface $X_k$.
Its components will be marked with indices outside the parentheses to
distinguish them from the number of the surface. We have
\be\label{g11=0}
(g_k)_{11}=-\frac{1}{2}\tr(\p X_k\c\p X_k)=\frac{1}{2}\tr([\p
P_k,P_k]\c [\p P_k,P_k])=-\frac{1}{2}\tr(\p P_k\c\p P_k)=0,
\ee
where we have successively applied the definition of the metric tensor,
the definition of $X_k$ given in \eqref{X}, property \eqref{exch-P} and
property \eqref{dPdP}. The vanishing of $(g_k)_{22}$ follows from the
Hermitian conjugate of \eqref{g11=0}.
\item
The nonzero off-diagonal element $(g_k)_{12}=(g_k)_{21}$ is equal
to
\be\label{g12}
(g_k)_{12}=-\frac{1}{2}\tr(\p X_k\c\bp X_k)=-\frac{1}{2}\tr([\p
P_k,P_k]\c [\bp P_k,P_k])=\frac{1}{2}\tr(\p P_k\c\bp P_k).
\ee
Thus the $1^\mathrm{st}$ fundamental form reduces to
\be\label{first}
I_k=\tr(\p P_k\c\bp P_k)\,d\xi d\bar{\xi}.
\ee
The second form
\be
\label{second} II_k=(\p^2 X_k-(\Gamma_k)^1_{11}\p
X_k)d\xi^2+2\p\bp X_k d\xi d\bar{\xi}+(\bp^2
X_k-(\Gamma_k)^2_{22}\bp X_k)d\bar{\xi}^2,
\ee
is easy to find when we determine the Christoffel symbols
$(\Gamma_k)^1_{11}$ and $(\Gamma_k)^2_{22}$. These are the only
nonzero components of the $\Gamma$. We have from \eqref{g12}
\be\label{Chris}
(\Gamma_k)^1_{11}=\p\ln{(g_k)_{12}},\quad
(\Gamma_k)^2_{22}=\bp\ln{(g_k)_{12}}.
\ee
Using \eqref{X} and the E-L equations \eqref{cons-law} together with
\eqref{Chris}, we can write \eqref{second} as
\be
\begin{split}
II_k&= -\tr(\p P_k\c\bp P_k)\,\p\frac{[\p P,P]}{\tr(\p P_k\c\bp
P_k)}d\xi^2+2[\bp P,\p P]d\xi d\bar{\xi}\\&+\tr(\p P_k\c\bp
P_k)\,\bp\frac{[\bp P,P]}{\tr(\p P_k\c\bp P_k)}d\bar{\xi}^2
\end{split}
\ee
Examples of the metric for surfaces induced by Veronese solutions
of the E-L equations \eqref{E-L} are given in \cite{GG1}.
\end{enumerate}

\section{Projectors and the spectral problem}
The spectral problem is closely related to the immersion functions of
the surfaces. The relation between the wave functions and the
immersion functions is given by the Sym-Tafel formula
\cite{Sym1,Sym2,Sym3,Taf}, and they are also related by their asymptotic
properties. These aspects of the theory were discussed in \cite{GG1}.
In this section we concentrate on the consequences of their
representation in terms of projectors.

Similarly to the surfaces, the wave functions of the spectral problem
can also be expressed in terms of the projectors. The spectral
problem found by Zakharov and Mikhailov \cite{MikZ} reads
\be\label{spectral}
\p \Phi_k=\frac{2}{1+\la}[\p P_k,P_k]\Phi_k , \qquad \bp
\Phi_k=\frac{2}{1-\la}[\bp P_k,P_k]\Phi_k, \qquad k=0,1,\ldots,N-1,
\ee
where $\la\in\mathbb{C}$ is the spectral parameter and the wave
functions are given by \cite{DHZ}
 \be\label{PhifromP}
\Phi_k=\mathbb{I}+\frac{4\la}{(1-\la)^2}\sum\limits_{j=0}^{k-1}P_j-\frac{2}{1-\la}P_k,
\ee\be
{\Phi_k}^{-1}=\mathbb{I}-\frac{4\la}{(1+\la)^2}\sum\limits_{j=0}^{k-1}P_j-\frac{2}{1+\la}P_k.
\ee
This in turn yields the projectors $P_k$ in terms of the wave
functions \cite{GG1}
\be\label{PfromPhi}
P_k=(1/4)\left[2(1+\la^2)\mathbb{I}-(1-\la)^2\Phi_k-(1+\la)^2\Phi_k^{-1}\right].
\ee
The projective property may be represented in terms of $\Phi_k$ as a
factorisable $4^\mathrm{th}$ degree expression with one double
(squared) factor, resembling the corresponding equation for the
surfaces $X_k$, namely
\be
P_k^2-P_k=(1/16)\Phi_k^{-2}(\mathbb{I}-\Phi_k)\left[(1+\la)^2
-(1-\la)^2\Phi_k\right]\left[(1+\la)-(1-\la)\Phi_k\right]^2=0.
\ee
It may be interpreted in the same way as the equivalent relation for
the surfaces \eqref{3-deg-cond}. We can also obtain a $3^\mathrm{rd}$
degree equation in which all of the linear factors are of the
$1^\mathrm{st}$ degree. This may be performed in a way similar to the
derivation of \eqref{3-deg-cond}, i.e. by multiplying
\eqref{PfromPhi} by $\Phi_k$ and applying the orthogonality of $P_k$
to $P_0\ldots P_{k-1}$.

\section{Concluding remarks}
The description of $\cpn$ models in terms of orthogonal projection
operators has a few advantages compared with their description in
terms of vectors. It is natural, and the picture which it provides is
clear. At the same time it need not be more difficult than that in
terms of vectors, provided that we know a few identities of the
projector algebra and analysis (such as those listed in Section 2).
The construction of an orthogonal basis in terms of projectors is
straightforward. Also the principal conditions required for the
consistency of the model are easy to prove.

The technique presented above for constructing an increasing number
of surfaces associated with $CP^{N-1}$ sigma models on Euclidean
spaces can lead to a detailed analytical description of the surfaces
in question. This description provides us with effective tools for
finding surfaces without invoking any additional considerations,
proceeding directly from the given $CP^{N-1}$ sigma model equations
(1.9).

In the next stage of this research, it would be worthwhile to extend
the presented approach to more general sigma models based on
Grassmannian manifolds, i.e. the homogeneous spaces
\begin{equation}
G(m,n)=\frac{SU(N)}{S\left(U(m)\times U(n)\right)},\quad N=m+n.
\label{Grass}
\end{equation}
Grassmannian sigma models are a generalization of $CP^{N-1}$ sigma
models. Their important common property is that the Euler-Lagrange
equations can best be written in terms of projectors. They share a
lot of properties like an infinite number of local and/or nonlocal
conserved quantities, infinite-dimensional symmetry algebras,
Hamiltonian structures, complete integrability, the existence of
multisoliton solutions, etc. The investigation of surfaces for this
case can lead to different classes and much more diverse types of
surfaces than the ones discussed in this paper. The geometrical
aspects of such surfaces will be described in more detail in a future
work.

\ack 
A.M.G.'s work was supported by a research grant from NSERC of Canada.
He also wishes to thank the Theoretical Physics Department of the
Andrzej Soltan Institute for Nuclear Studies for their hospitality.
P.P.G. wishes to acknowledge and thank the Mathematical Physics
Laboratory of the Centre de Recherches Math\'{e}matiques for their
hospitality during his visit to the Universit{\'e} de Montr{\'e}al,
where work on this topic was begun.

\appendix

\section{Proofs of consistency properties}
Here we use the properties (i--x) of the projection operators mapping
onto one-dimensional subspaces in order to prove several properties
required for the consistency of the description of the $\cpn$ model in
terms of projection operators.
\begin{enumerate}
\item
The projective property of the $P$ ``promoted'' by the creation
operator \eqref{Ppm} has been proven in \cite{GG1}. However we may
obtain it immediately by using the property \eqref{tr-P}
\be\label{Pi+Pi+}
\mathbf{\Pi_+}(P)\c \mathbf{\Pi_+}(P)=\frac{\p P\c P\c \bp P\c \p
P\c P\c \bp P}{[\tr(\p P\c P\c\bp P)]^2}=\frac{\p P\c P\c \bp
P}{[\tr(\p P\c P\c\bp P)]}=\mathbf{\Pi_+}(P),
\ee
where the property \eqref{tr-P} has been applied to transform the
numerator in \eqref{Pi+Pi+} according to
\be\label{numerat}
(\p P\c P\c \bp P)\c (\p P\c P\c \bp P)=\tr(\p P\c P\c \bp P)\,\p P\c
P\c \bp P.
\ee
A by-product of the proof is a demonstration that the trace of $\p
P\c P\c \bp P$ vanishes iff the whole matrix vanishes. Namely, if the
trace vanishes then the r.h.s. of \eqref{numerat} is zero, but it is
a square of a Hermitian matrix $\p P\c P\c \bp P$. Hence it vanishes
iff the matrix vanishes.

This means that the construction of $\mathbf{\Pi_+}(P)$ from $P$
\eqref{Pi+} is correct and always possible, except for the cases in
which $\p P\c P\c \bp P$ vanishes. In this case $\mathbf{\Pi_+}(P)=0$
is indeterminate.

Mutatis mutandis we may prove the projective property and correctness
of the construction for $\mathbf{\Pi_-}(P)$.

The orthogonality of the projectors $\mathbf{\Pi_+}(P)$ and
$\mathbf{\Pi_-}(P)$ follows from the fact that they are Hermitian
(which may be checked in a straightforward way).
\item
In order to make the model consistent, we should have
\be
\mathbf{\Pi_-}\left(\mathbf{\Pi_+}(P)\right)=P\quad \mathrm{and}
~~\mathbf{\Pi_+}\left(\mathbf{\Pi_-}(P)\right)=P,
\ee
provided that the action of the creation operator (first case) or the
annihilation operator (second case) can be executed. For shorthand
notation we introduce
\be
P_{+1}=\mathbf{\Pi_+}(P); \quad P_{-1}=\mathbf{\Pi_-}(P)
\ee
According to the definition of $\mathbf{\Pi_\pm}$,we have to prove
that $\bp P_{+1}\c P_{+1}\c\p P_{+1}/\tr(\bp P_{+1}\c P_{+1}\c\p
P_{+1})=P$, provided that $P_{+1}\ne 0$.

To assess $\bp P_{+1}\c P_{+1}\c\p P_{+1}$ and its trace, note that
we may double $P_{+1}$ in this expression, due to its previously
proven projective property.
\be\label{P+1P+1}
\bp P_{+1}\c P_{+1}\c\p P_{+1}=\bp P_{+1}\c P_{+1}\c P_{+1}\c\p
P_{+1}
\ee
Note also that the second half of the r.h.s. in equation \eqref{P+1P+1} is the
Hermitian conjugate of its first half. Consider the first half. We
have
\be
\begin{split}
\bp P_{+1}\c P_{+1}&=\bp\frac{\p P\c P\c\bp P}{\tr(\p P\c P\c\bp
P)}\c\frac{\p P\c P\c\bp P}{\tr(\p P\c P\c\bp P)}\\&=\frac{(P\c\bp\p
P\c\bp P+\p P\c\bp P\c\bp P)\p P\c P\c\bp P}{\left[\tr(\p P\c P\c\bp
P)\right]^2}.
\end{split}
\ee
We apply properties \eqref{even-P} and \eqref{tr-P} in order to
replace some of the factors in the numerator by the appropriate
traces, then we use the property \eqref{factor-P} to factor those
traces. The invariance of traces under cyclic permutations of factors
allows us to obtain (after cancellation of the common factor)
\be
\bp P_{+1}\c P_{+1}=\frac{\left[\tr(\bp\p P\c P)+\tr(\p P\c \bp P\c
P)\right] P\c\bp P}{\tr(\p P\c P\c\bp P)}
\ee
Using the property \eqref{2-d deriv} and then the exchange property
\eqref{exch-P}, we further get
\be\label{P+1 from P}
\bp P_{+1}\c P_{+1}=\frac{\left[-\tr(\p P\c\bp P)+\tr(\p P\c \bp P\c
P)\right] P\c\bp P}{\tr(\p P\c P\c\bp P)}=-P\c\bp P
\ee
Hence, combining the above result with its Hermitian conjugate, we
finally get
\be
\mathbf{\Pi_-}\left(\mathbf{\Pi_+}(P)\right)=\frac{P\c\bp P\c\p P\c
P}{\tr(P\c\bp P\c\p P\c P)}=P
\ee
due to property \eqref{tr-P}. \textit{Q.E.D.}
\item
We now prove, by a different method than that of \cite{GY1}, the
equation \eqref{XfromP}, expressing $X_k$ as a sum of projectors. An
equation obtained in the proof will also be used to demonstrate
another property necessary for the consistency of the model: if $P_m$
satisfies the E-L equation \eqref{cons-law}, then
$P_{m+1}=\mathbf{\Pi_+}(P_m)$ also does.

\textit{Proof:} The surfaces $X_k$ are defined by \eqref{X} up to a
constant matrix (whose diagonal elements are uniquely determined by
the condition that the traces of $X_k$ vanish). Hence it is
sufficient to prove that
\be\label{Xk_from_Pk-a}
\p X_k=-i\p P_k-2i\sum\limits_{j=0}^{k-1}\p P_j,
\ee
\be\label{Xk_from_Pk-b}
\bp X_k=-i\bp P_k-2i\sum\limits_{j=0}^{k-1}\bp P_j,
\ee
or equivalently
\be\label{commut from Pk-a}
[\p P_k,P_k]=\p P_k+2\sum\limits_{j=0}^{k-1}\p P_j,
\ee
\be\label{commut from Pk-b}
[\bp P_k,P_k]=-\bp P_k-2\sum\limits_{j=0}^{k-1}\bp P_j.
\ee
This thesis will be proven by induction. For $k=0$, equations (\ref{commut
from Pk-a}) and (\ref{commut from Pk-b}) reduce to
\be\label{commut0 from P0}
\p P_0 P_0-P_0\p P_0=\p P_0\quad \text{and }~\bp P_0 P_0-P_0\bp
P_0=-\bp P_0
\ee
which by property \eqref{exch-P} are equivalent to
\be
P_0\c\p P_0=0 \quad\text{and }~\bp P_0\c P_0=0
\ee
where $P_0$ maps onto a direction of a holomorphic function. Let $z$
be a unit vector in that direction. Then $P_0=z\otimes z$ and $z$
depends only on $\xi$, while $z^{\dagger}$ depends only on $\bar{\xi}$. Hence
\be
P_0\c\p P_0=z\otimes z^{\dagger}\c\p z\otimes
z^{\dagger}=(z^{\dagger}\c\p z)z\otimes z^{\dagger}=0
\ee
since $z^{\dagger}\c\p z=\p (z^{\dagger}\c z)=0$ for any holomorphic
unit vector $z$. The second half of \eqref{commut0 from P0} is the
Hermitian conjugate of the first half.

Let (\ref{commut from Pk-a} and \ref{commut from Pk-b}) now hold
for $k=m\ge 0$. By property \eqref{exch-P} we have
\be
\bp P_{m+1}\c P_{m+1}-P_{m+1}\c\bp P_{m+1}=2\p P_{m+1}\c P_{m+1}-\bp
P_{m+1}
\ee
We may replace $2\,\bp P_{m+1}\c P_{m+1}$ by $-2\,P_m\bp P_m$ using
\eqref{P+1 from P}. Applying property \eqref{exch-P} to one of these
two $P_m\bp P_m$ we get
\be\label{commut+1from commut}
[\bp P_{m+1},P_{m+1}]=-\bp P_{m+1}+[\bp P_m,P_m]-\bp P_m
\ee
On the basis of the induction hypothesis, \eqref{commut+1from commut}
turns into
\be
[\bp P_{m+1},P_{m+1}]=-\bp P_{m+1}-2\sum\limits_{j=0}^m\bp P_j,
\ee
which is exactly the second part of the thesis, i.e. \eqref{commut
from Pk-b}. The first part of the thesis: \eqref{commut from Pk-a} is
its Hermitian conjugate. \textit{Q.E.D.}

From the intermediate result \eqref{commut+1from commut} we obtain the
connection between the E-L equations \eqref{cons-law} for $P_m$ and
those for $P_{m+1}$. Applying the $\p$ derivative to
\eqref{commut+1from commut}, the $\bp$ derivative to its Hermitian
conjugate, and subtracting the results of the differentiation from
each other we obtain
\be
\p\,[\bp P_{m+1},P_{m+1}]+\bp\,[\p P_{m+1},P_{m+1}]=\p\,[\bp
P_m,P_m]+\bp\,[\p P_m,P_m],
\ee
whence $P_{m+1}$ satisfies the E-L equations \eqref{cons-law} iff
$P_m$ does, provided that the construction of $P_{m+1}$ from $P_m$ is
possible. This result may also be used in the opposite direction:
when the construction of $P_{m-1}$ from $P_m$ is possible, then
$P_{m-1}$ satisfies the E-L equations iff $P_m$ does. This is another
criterion of consistency of the model.

Another by-product of the proof is the demonstration of property (x)
\eqref{dPdP} (from which it follows that the metric tensor $g_k$ of
the surfaces $X_k$ has zeros on the diagonal
$(g_k)_{11}=(g_k)_{22}=0$). Indeed \eqref{dPdP} obviously holds for
$k=0$, when $P=z\otimes z^\dagger$. If it holds for $k=m$ then,
writing \eqref{commut+1from commut} as $[\bp P_{m+1},P_{m+1}]+\bp
P_{m+1}=[\bp P_m,P_m]-\bp P_m$ and squaring both sides, we obtain
(using property \eqref{exch-P} and the invariance of traces under cyclic
permutations)
\be
\tr(\bp P_{m+1}\c\bp P_{m+1})=\tr(\bp P_{m}\c\bp P_{m}),
\ee
which yields the thesis for all $k$ by induction.
\end{enumerate}
\section*{References}

\begin{footnotesize}

\end{footnotesize}

\end{document}